\documentclass[12pt]{iopart}

\usepackage{iopams}
\usepackage{amssymb}
\usepackage{graphicx}
\eqnobysec

\newcommand{\re}{\mathrm{Re}\,}
\newcommand{\im}{\mathrm{Im}\,}

\newcommand{\sign}{\mathrm{sign}\,}

\begin{document}

\title[On the Burgers vector of a wave dislocation]{On the Burgers vector of a wave dislocation}

\author{Mark R Dennis}

\address{H H Wills Physics Laboratory, Tyndall Avenue, Bristol BS8 1TL, UK}

\begin{abstract}
Following Nye and Berry's analogy with crystal dislocations, an approach to the Burgers vector of a wave dislocation (phase singularity, optical vortex) is proposed.
It is defined to be a regularized phase gradient evaluated at the phase singularity, and is computed explicitly.
The screw component of this vector is naturally related to the helicoidal twisting of wavefronts along a vortex line, and is related to the helicity of the phase gradient.
The edge component is related to the nearby current flow (defined by the phase gradient) perpendicular to the vortex, and the distribution of this component is found numerically for random two-dimensional monochromatic waves.
\end{abstract}


\section{Introduction}\label{sec:int}

In their study of phase singularities as a general phenomenon in wave interference and diffraction, Nye and Berry called them {\em wave dislocations} \cite{nb:1974b34}.
Other names used for these objects include optical vortices (as they are circulations of the flow associated with the waves), and nodal points in two dimensions, and nodal lines in three, as they are also the zero loci of the wave intensity.

The term `dislocation' is more established for defects in crystals, that is, imperfections in a regular periodic lattice geometry, which also occurs at points in two dimensions and on lines in three, and were originally proposed as a general feature of real crystal lattices in the 1950s \cite{frank:1951dislocations,read:1953dislocations} \footnote{It is interesting to note that around the same period, special examples of optical vortices were first noticed and studied \cite{dop:2009pio}.}.
Since then, the study of geometric topological defects has played a key role in condensed matter physics and beyond (e.g.~\cite{mermin:1979defects,vachaspati:1998cosmosandlab}).

These defects are topological because their presence signifies a breakdown in a key geometric descriptor of a spatially extended field (the order parameter), and a net change in that parameter in the vicinity of the defect, for instance on taking a circuit around it.
For phase singularities, this is the strength: the phase change, quantized in units of $2\pi,$ around the node.
For nodal points in two dimensions, this gives a signed integer (topological charge), whereas for lines in three dimensions, the sign endows the line with an orientation (topological current): a natural direction for the tangent at each point along the nodal line.

\begin{figure}
\begin{center}
\includegraphics*[width=10cm]{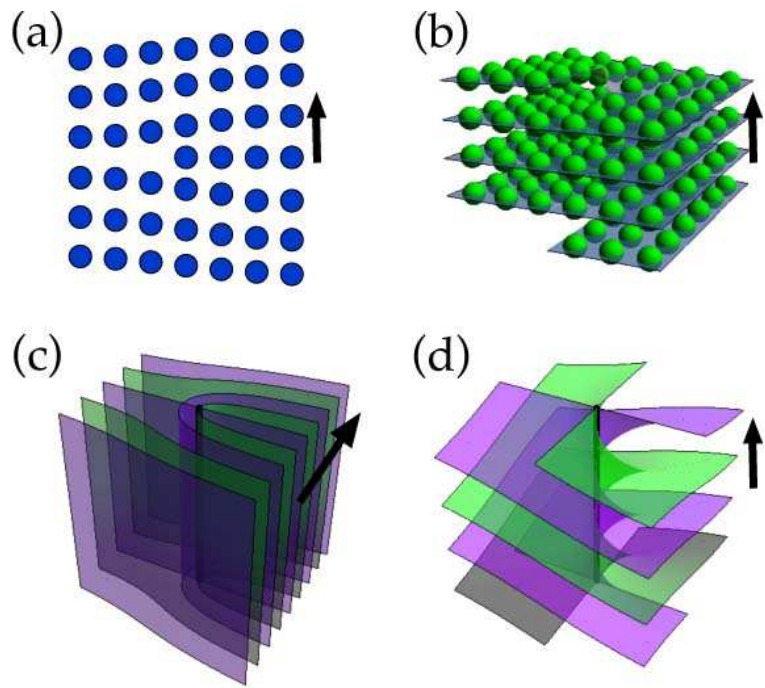}
\end{center}
    \caption{Wave and crystal dislocations. (a) edge and (b) screw dislocations in crystal lattices, with arrows showing the Burgers vector directions.
    (c) and (d) show wavefronts (zero contour surfaces of real and imaginary parts) of the edge (c) and screw (d) dislocated waves described by Eq.~(\ref{eq:wavedislocation}), with propagation directions $\bi{k}$ given by the arrows.
     }
    \label{fig:dislocation}
\end{figure}

A dislocation in a crystal lattice, on the other hand, are defined by its {\em Burgers vector} \cite{frank:1951dislocations}: the vector by which a circuit around the dislocation would fail to close in a corresponding circuit in a perfect lattice.
It is thus a vector with integer components, the two most distinctive being an edge dislocation, whose Burgers vector is perpendicular to the direction of the dislocation line (all dislocations in 2-dimensional lattices are of this type), and a screw dislocation, whose Burgers vector is parallel to the line's direction.
Examples of these two kinds are shown in Figure \ref{fig:dislocation}(a) and (b).

Crystal and wave dislocations are described by different mathematical objects since the fields of which they are defects have different physical nature -- a wave field, denoted here by $\psi,$ is a smooth, complex function of position, and a crystal is a discrete lattice generated by specific lattice vectors.
Thus, does it make sense to define a Burgers vector for a wave dislocation?

Nye and Berry's choice of nomenclature was motivated by specific, simple examples where dislocated waves have similarities to their crystal counterparts.
Specifically, they considered a time-independent wave function of the form
\begin{equation}
   \psi_{\bi{k}}(\bi{r}) = (x + \rmi y) \exp(\rmi \bi{k}\cdot \bi{r}),
   \label{eq:wavedislocation}
\end{equation}
with a node where $x = y = 0$ (giving a point at the origin in two dimensions, and a line along the $z$-axis in three).
This is modulated by a plane wave with wavevector $\bi{k}.$
With $\bi{k}$ in the $x,y$-plane (say the $y$-direction), the pattern of constant phase lines resembles a crystal edge dislocation.
In three dimensions, if $\bi{k}$ is parallel to $\hat{\bi{z}},$ the surfaces of constant phase are helicoids and resemble a screw.
Examples of these surfaces of constant phase in three dimensions are shown in Figure \ref{fig:dislocation} (c,d).
Helical screw dislocations occur on the axis of optical beams carrying orbital angular momentum, such as Laguerre-Gauss modes \cite{bvs:1990screw,apb:1999oam}.
Mixed edge-screw dislocations occur as intermediates.

However, general wavefields, typified by superpositions of plane waves propagating in different directions, do not in general have an overall carrier wavevector, although they do have a complicated configuration of nodal points or lines.
Nevertheless, at points where the phase is not singular, it is natural to define the local wavevector as the local phase gradient \cite{bcluw:1980b96,dhc:2008superoscillation,berry:2009currents}.
In quantum waves, this definition is related to the concept of weak measurements (i.e.~local rather than global averages) \cite{ar:2005quantum}.
Here, I will consider a general definition for the Burgers vector of a wave dislocation, based on a regularized phase gradient at the phase singularity.

This work extends my previous approach \cite{dennis:2004twirl} to the local phase structure of a wave dislocation.
There, I considered several different measures for the rate of twist of the phase helicoids around a general dislocation line.
This twist geometry is quite complicated, since the intensity contours in the plane transverse to a dislocation line are ellipses, with associated nonuniformity in phase \cite{dennis:2001local,efv:2004finestructure,roux:2004noncanonical}, and generically this anisotropy ellipse rotates -- `twirls' -- along the singular line, resulting in the different phase helicoids twisting at different rates.
This was used to define an average rate of rotation, the `ellipse-defined twist', of the phase core along a singularity line (\cite{dennis:2004twirl} Eq.~(2.8)),
\begin{equation}
   Tw_{\rm{ell}}  = \frac{\rm{Re}\{ \bi{T} \cdot(\nabla \psi^{\ast} \times \nabla
   \psi') + \rmi \nabla \psi^{\ast}\cdot \nabla \psi'\}}{|\nabla \psi|^2
   +|\nabla \psi^* \times \nabla \psi|},
   \label{eq:twell}
\end{equation}
where $\bullet' = \bi{t}\cdot \nabla \bullet,$ and $\bi{t}$ is the tangent vector of the dislocation line (whose direction is provided by the topological orientation), parallel to $\rm{Im} \nabla \psi^* \times \nabla \psi$ \cite{bd:2000b321}.
As a measure of screwness, this quantity has certain attractive properties, although its derivation was somewhat oblique: it is a pure derivative of a certain angle, implying that it integrates to an integer times $2\pi$ around a vortex loop (the screw number of the dislocation), and, in random waves, its fluctuations were smaller than the other measures considered.
I will show here that $Tw_{\rm{ell}}$ arises naturally as the tangential screw component of the wave dislocation's Burgers vector, and find the corresponding transverse part.
Other measures considered in Ref.~\cite{dennis:2004twirl}, such as the azimuth-averaged twist $Tw_{\phi}$ (the average rate of rotation of all phase surfaces along the singularity), and the screwness $\sigma$ (the rate of change of phase for fixed azimuth), appear as the tangential components of alternative definitions of the Burgers vector.

Nye proposed \cite{nye:1981motion} a similar definition of the Burgers vector for dislocations in time-dependent waves (based on a generalization of $\psi_{\bi{k}}$), although it is difficult to generalize his definition to time-independent waves.
In particular, his definition is based around a specific coordinate system attached to each point on the singularity, based on the tangent to the singularity and its perpendicular direction of motion. 
Nye's Burgers vector is then the limit of the phase gradient on approaching the singularity from these specific orthogonal directions; when the dislocation is not moving, these directions cannot be defined.
The Burgers vector defined in this paper is based on a local average of the phase gradient at points in a neighbourhood perpendicular to the dislocation line; the sense of circulation around the phase vortex cancels out in this average, leaving the residual phase gradient which defines the Burgers vector.
The waves considered here will be time independent and monochromatic (although these assumptions are not completely necessary).

The structure of this paper is as follows.
In the following section, basic notions associated with phase gradient and phase singularities will be described, along with the quantum-style Dirac notation required to write the Burgers vector.
The derivation and basic description of the Burgers vector is given in Section \ref{sec:burgers}, culminating in its definition in Eq.~(\ref{eq:burgers}).
The details of the calculation here is given in the Appendix.
Readers interested only in the consequences of the definition may omit these sections.
The direction of the Burgers vector in the direction of the dislocation tangent -- the screw component -- determines the helicoidal twisting of the phase surfaces along the line, and this is related to the notion of helicity in fluid flows in Section \ref{sec:screw}.
The remainder, a vector perpendicular to the singularity line, defines the edge component, is discussed in Section \ref{sec:edge}.
In particular, the statistical distribution of the edge component in isotropic 2-dimensional monochromatic waves are explored numerically.

\section{Local wave dislocation geometry}
\label{sec:local}

The following discussion applies to a general complex scalar wavefunction 
\begin{equation}
   \psi = \rho \exp(\rmi \chi),
   \label{eq:psidef}
\end{equation}
with intensity $\rho^2,$ and phase $\chi.$
Explicit dependence on position (either in two or three dimensions) will usually be omitted.
Wave dislocations are loci where $\psi = 0;$ on these points (2D) or lines (3D), the intensity vanishes quadratically, and $\chi$ is undefined.
As vortices, the dislocations are also circulations of the current associated with the wave,
\begin{equation}
   \bi{j} = \im \psi^* \nabla \psi.
   \label{eq:jdef}
\end{equation}
The wave Burgers vector, defined in the next section, will be expressed in terms of various derivatives of the field close to the dislocation, which are described in this section.
Similar discussions of this geometry may also be found in Refs.~\cite{dop:2009pio,dennis:2004twirl,bd:2000b321,dennis:thesis}.

The phase is a well-defined function at typical points off the nodes; its gradient may be written
\begin{equation}
   \nabla \chi = \im \nabla \log \psi = \im \frac{\nabla \psi}{\psi} = \frac{\bi{j}}{\rho^2}.
   \label{eq:gradchidef}
\end{equation}
$\nabla \chi$ is the local wave momentum at the point evaluated, i.e.~the local wavevector \cite{bcluw:1980b96,dhc:2008superoscillation}.
For waves described by a vector field (such as the electric vector for polarized light), the natural definition for local propagation direction \cite{nye:1991frank} can be written as a local wavevector in an analogous way; these quantities for non-singular points are explored in more detail by Berry \cite{berry:2009currents}.
When $\psi$ is zero, however, $\nabla \chi$ is singular, since $\psi$ and $\rho$ are zero -- this definition for a local wavevector fails.

As described in the introduction, the topological strength $s$ of a phase singularity is determined by a line integral of phase on a closed circuit $C$ around the singularity.
Using Stokes' theorem, 
\begin{equation}
   s = \frac{1}{2\pi} \oint_C \rmd \bi{r}\cdot \nabla \chi = \frac{1}{2\pi} \int_{S} \rmd^2\bi{r}\cdot (\nabla \times \nabla \chi),
   \label{eq:curljust1}
\end{equation}
where $S$ is the surface or area enclosed by $C,$ and $s$ is the integer representing the net number of circuits of phase executed with respect to $C;$ it will be assumed throughout this paper that the singularity has unit strength, i.e.~$s = \pm 1.$
Since $\nabla \chi$ is a well defined gradient off the singularity, its curl vanishes there, and the only contribution to the integral in (\ref{eq:curljust1}) can be a $\delta$-function at the node,
\begin{equation}
   \nabla \times \nabla \chi = \pi \delta^2(\psi) (\nabla \times \bi{j}).
   \label{eq:curljust2}
\end{equation}
(This may be more rigorously justified by regularizing $\nabla \chi$ at the singularity.)
Equation (\ref{eq:curljust2}) shows that the wave's vorticity -- the curl of the flow velocity $\nabla \chi$ -- is concentrated on the vortex lines with weighting
\begin{equation}
   \bi{\Omega} = \frac{1}{2} \nabla \times \bi{j} = \frac{-\rmi}{2} \nabla \psi^* \times \nabla \psi.
   \label{eq:Omegadef}
\end{equation}
$\bi{\Omega}$ alone is often referred to as the vorticity, although as the curl of $\nabla \chi$ (rather than $\bi{j}$) it is only nonzero on the nodal lines.
In three dimensions, $\bi{\Omega}$ points in the vortex line tangent direction $\bi{t}$ in the direction of topological current; in two (denoted by $x,y$), its scalar sign ($\bi{\Omega}$ dotted with the unit $z$-vector) defines the topological charge sign.

It is convenient to choose local cartesian coordinates defined by this direction.
With the origin on the singular line, with tangent in the $z$-direction (parallel or antiparallel to the topological current), the 2-dimensional vorticity may be written
\begin{equation}
   \bi{\Omega}\cdot\hat{\bi{z}} = \frac{-\rmi}{2} (\psi_x^* \psi_y - \psi_y^* \psi_x) = \im \frac{1}{2} \nabla \psi^* \cdot \boldsymbol{\sigma}_3 \cdot \nabla \psi.
   \label{eq:S3def}
\end{equation}
Acting between the gradient vectors in the final equality is the Pauli matrix $\boldsymbol{\sigma}_3 = \left(\begin{array}{cc} 0 & -\rmi \\ \rmi & 0 \end{array}\right).$
Similar inner products with respect to the other Pauli matrices (noting that the labels are permuted from their usual place, as they are here applied to vectors in a cartesian basis, rather than a polar basis), for $j = 0,1,2,3,$ give
\begin{equation}
   S_j \equiv \nabla \psi^* \cdot \hat{\sigma}_j \cdot \nabla \psi.
   \label{eq:stokesdef}
\end{equation}
These Stokes-like parameters describe the local anisotropy of the gradient vector $\nabla \psi$ \cite{dennis:2004twirl,efv:2004finestructure,roux:2004noncanonical,dennis:thesis}.
Thus $S_0 = |\nabla \psi|^2,$ and $S_3 = 2 \bi{\Omega}\cdot \hat{\bi{z}},$ whose sign gives the singularity strength.
The orientation angle of the anisotropy ellipse defined by $\nabla \psi$ is given by $\frac{1}{2}\arg(S_1 + \rmi S_2).$

The anisotropy ellipse describes the elliptical contours of low intensity, and related squeezing of phase contours around the vortex; since the flow of current $\bi{j} = \rho^2 \nabla \chi$ is a perfect circulation, these two effects cancel \cite{bd:2000b321,dennis:2001local}.
The major axis of the $\nabla \psi$ ellipse corresponds to the minor axis of the local intensity ellipse, and vice versa.

\section{The Burgers vector as a regularized phase gradient}
\label{sec:burgers}

In this section, the Burgers vector will be defined following the definitions and notation of the previous section.
With coordinates chosen so the dislocation is at the origin (with direction in the $z$-direction for three-dimensional waves, and the $z = 0$ plane is considered), the phase gradient $\nabla \chi$ is singular as the cylindrical radius $R = \sqrt{x^2+y^2}$ approaches 0.
Around the dislocation $\psi = 0,$ at which the field gradient is $\nabla\psi_0,$ the phase gradient $\nabla \chi = \im \nabla \psi/\psi$ may be expanded in powers of $R,$
\begin{equation}
   \nabla \chi = \frac{1}{R} \im\left(\frac{\nabla \psi_0}{\bi{u}\cdot\nabla\psi_0}\right) 
   + \im\left[\frac{\mathbf{\Psi}\cdot\bi{u}}{\bi{u}\cdot\nabla\psi_0} - \frac{\nabla \psi_0}{2} \frac{\bi{u}\cdot\mathbf{\Psi}\cdot\bi{u}}{(\bi{u}\cdot\nabla\psi_0)^2}\right]
   +O(R),
   \label{eq:gradchiexpand}
\end{equation}
where $\bi{u}$ denotes the unit radius vector $\bi{u} = (x,y,0)/R = (\cos \phi, \sin \phi,0)$ and $\mathbf{\Psi}$ is the hessian matrix of second derivatives, $\Psi_{ij} = \partial_i \partial_j \psi,$ $i,j = x,y,z.$
The first term of (\ref{eq:gradchiexpand}) is proportional to $R^{-1},$ accounting for the singularity.
The numerator of this term proportional to the direction field for the current near the singularity, $\bi{j} = R \, \im( (\nabla \psi \cdot \bi{u}) \nabla \psi^*) + O(R^2),$ known to be a perfect circulation \cite{bd:2000b321,dennis:2001local}.

The second term is independent of $R,$ although dependent on unit direction $\bi{u}.$
Its two components (within $[ \; ]$ brackets) arise from the next-to-leading order terms in the expansion of $\nabla \psi/\psi,$ and depends linearly ion the second derivatives $\mathbf{\Psi}.$  
Only the first of these has a component in the $z$-direction, as the other is proportional to $\nabla \psi_0,$ which is by definition orthogonal to $\hat{\bi{z}}.$
The expression in $[ \; ]$ accounts for $\nabla \chi$ at the singularity (up to terms of order $R$), aside from the singular circulation in $R^{-1}.$

Evaluating this $R$-independent term for the dislocated waves (\ref{eq:wavedislocation}) indeed gives the carrier wavevector $\bi{k};$ this in fact is also the case for the more general form $\psi = ((\bi{X} + \rmi \bi{Y})\cdot\bi{r}) \exp(\rmi \bi{k}\cdot\bi{r}),$ for $\bi{X}, \bi{Y}$ any pair of linearly independent vectors in the $x,y$-plane.
For these simple cases, the local wavevector is simply the non-circulatory part of $\nabla \chi$ in the limit $R \to 0,$ and is therefore a natural definition of a local wavevector.
For more general waves, however, this is dependent on $\bi{u},$ and so this limit must be approached democratically from all directions.

The Burgers vector $\bi{b}$ for a wave dislocation in a typical field, therefore, is defined to be the integral of the phase gradient (\ref{eq:gradchiexpand}) over a small circular disk in the $xy$-plane centred on the singularity, in the limit of the disk radius $d \to 0,$
\begin{equation}
   \bi{b} = \lim_{d\to 0} \frac{1}{\pi d^2}\int_{\mathrm{disk}}\rmd^2\bi{r} \, \nabla \chi = \frac{1}{2\pi}\oint \rmd\phi \, \nabla \chi.
   \label{eq:bintdef}
\end{equation}
The singular circulatory term cancels (since $\psi$ is averaged before $R \to 0$), so $\bi{b}$ is determined purely by the term in square brackets $[ \; ]$ in (\ref{eq:gradchiexpand}).
This definition is more general, but similar, to Nye's definition of a Burgers vector in time-dependent waves \cite{nye:1981motion}.

These integrals, involving the ratios of linear and quadratic forms in $\bi{u},$ are straightforward to integrate using complex contour integration; this is described in \ref{sec:contour}.
The singular term in $R^{-1}$ integrates to zero, constituting the regularization.
Explicitly, the integral of the terms in $[ \; ]$ in (\ref{eq:gradchiexpand}) gives
\begin{equation}
   \bi{b} = \im\left[ \frac{\mathbf{\Psi}\cdot\bi{e}_s^*}{\bi{e}_s^* \cdot \nabla \psi_0} - \frac{\nabla\psi_0}{2} \frac{\bi{e}_s^*\cdot\mathbf{\Psi}\cdot\bi{e}_s^*}{(\bi{e}_s^* \cdot \nabla \psi_0)^2}\right],
   \label{eq:burgers}
\end{equation}
where $\bi{e}_s$ is the isotropic circular vector with the same sense $s$ as the singularity (from (\ref{eq:curljust1})),
\begin{equation}
   \bi{e}_s = \hat{\bi{x}} + s \rmi \hat{\bi{y}}.
   \label{eq:esdef}
\end{equation}
$\bi{e}_s$ as defined here may be multiplied by an arbitrary complex number (e.g.~normalizing it) without affecting $\bi{b}.$
The two terms of $\bi{b}$ arise directly from integrating the two terms in $[\;]$ in (\ref{eq:bintdef}).

Equation (\ref{eq:burgers}) for the Burgers vector $\bi{b},$ as a regularized phase gradient at the phase singularity, is the main result of this paper.
Geometrically, $\bi{b}$ is a measure of the local curvature of phase lines or surfaces local to a dislocation; precisely how $\bi{b}$ as defined is related to this curvature is not clear.
A short calculation of the $z$-component $\bi{b}_z$ gives $-Tw_{\rm{ell}},$ the negative of the twist defined in (\ref{eq:twell}).
The significance and meaning of this will be discussed in section \ref{sec:screw}.
The edge component perpendicular to the vortex, in the $xy$-plane, will be discussed in section \ref{sec:edge}.

The choice of regularizing by integrating over a disk centred on the singularity is natural, but not unique. 
Another obvious choice is to integrate over the anisotropy ellipse.
Effectively, this may be done by replacing $\bi{u}$ with $\re(\rme^{-\rmi s \theta} \nabla \psi_0)$ and integrating with respect to $\theta.$
This gives a phase-weighted Burgers vector similar to $\bi{b},$ but with $\bi{e}_s^*$ in (\ref{eq:burgers}) replaced by $\nabla\psi_0^*;$ the third component of this vector, along the direction of the singularity line, is minus the screwness $\sigma$ defined in Ref.~\cite{dennis:2004twirl} Eq.~(2.4).
Averaging with respect to the intensity anisotropy ellipse, whose axes are orthogonal to those of $\nabla \psi_0,$ is achieved by using the vector $\re(\rme^{-\rmi s \theta}(-\psi_{0y}\hat{\bi{x}} + \psi_{0x}\hat{\bi{y}})).$
The result is again the same as (\ref{eq:burgers}), but with with $\bi{e}_s^*$ replaced by $-\psi_{0y}^*\hat{\bi{x}} + \psi_{0x}^*\hat{\bi{y}},$ and the component along the singularity line is the negative of the azimuth-averaged twist $Tw_{\phi}$ of Ref.~\cite{dennis:2004twirl} Eq.~(2.3).
Each of these alternative definitions give the desired vector $\bi{k}$ for singularities of the form (\ref{eq:wavedislocation}), although they are different for more general waves.
It appears most natural to define the Burgers vector weighting all spatial directions around the singularity equally, as above in (\ref{eq:bintdef}), (\ref{eq:burgers}); in particular, the alternative definitions do not have the attractive property of being related to helicity, as described for $\bi{b}$ in the following section.

\section{The screw component}
\label{sec:screw}

The screw nature of a wave dislocation line in three dimensions, as the magnitude of the Burgers vector in the direction of the singularity line $\bi{t}\cdot \bi{b} = -Tw_{\rm{ell}},$ is a measure of the rate of rotation of the phase structure along the dislocation line.
The $-$ sign can be justified from the well-known fact that for a screw dislocation of the form (\ref{eq:wavedislocation}), the helicoid is left-handed when $\bi{\Omega} \cdot \bi{k} > 0,$ and right-handed otherwise; $Tw_{\rm{ell}}$ was defined in \cite{dennis:2004twirl} to be positive for a right-handed screw.

This quantity, integrated over a simply-connected volume $\mathcal{V}$ whose surface is not pierced by any singularities, it is reminiscent of the helicity $H$ of a fluid flow with velocity field $\bi{v},$
\begin{equation}
   H = \int_{\mathcal{V}} \rmd^3 \bi{r}\, \bi{v}\cdot\bi{\nabla \times \bi{v}}.
   \label{eq:helicity}
\end{equation}
Helicity is a topological invariant: provided the topology of the vortex lines does not change, $H$ is constant.
With $\bi{v} = \nabla \chi,$ the integrand is $2\pi \delta^2(\psi) \bi{\Omega}\cdot\bi{b},$ so the helicity of a scalar optical field is the integral of the screw component of the Burgers vector $-Tw_{\rm{ell}}$ along all of the closed loops contained in the volume $\mathcal{V}.$

Around each such loop, the phase pattern must rotate an integer number of times, which may be thought of as the topological linking number of the edges of any local phase ribbon centred on the singularity \cite{dennis:thesis,dennis:2004twirl,dh:2005calugareanu}.
For a planar loop, this linking number is simply the integral of $\bi{\Omega}\cdot\bi{b}$ around the loop, and if the loop is non-planar, it is the twist $Tw$ for the C{\u a}lug{\u a}reanu-White-Fuller relation $Lk = Tw + Wr$ (with linking $Lk$ and writhe $Wr$) \cite{dh:2005calugareanu,fuller:decomposition}.
The helicity $H$ is also related to this theorem in more general fluid mechanics \cite{mr:1992calugareanu}.

The significance of the previous discussion is that a closed dislocation loop with nonzero helicity -- for which there is a net integer number of screw rotations of the phase around the loop -- must be threaded by other dislocation lines \cite{bd:2001b332,ws:1983filaments2}.
Knotted (self-linked) dislocation lines have a helicity linking number given by their minimum crossing number \cite{ws:1983filaments3}. 
This result is a consequence of the continuity of the complex scalar field in three dimensions: the surfaces of constant phase must fill all space, only crossing each other on the singularity lines, and all phase surfaces must cross there.

A simple example of a wave (solving the Helmholtz equation) with such a threaded loop is \cite{bd:2001b332,dennis:2004twirl}
\begin{equation}
   \psi_{\rm{threaded}} = R^m \exp(\rmi m s \phi) \left( k(R^2 -a^2) + 2\rmi (1+m) z\right) \exp(\rmi k z),
   \label{eq:threaded}
\end{equation}
in cylindrical coordinates $(R,\phi,z),$ and integer $m >0.$ 
This field has a straight dislocation line of strength $m$ along the $z$-axis, with topological current parallel to $s\hat{\bi{z}}$, encircled by a nodal loop of radius $a$ in the $z = 0$ plane.
This loop has both screw and edge character.
The topology requires the integral of the screw component around the loop to be $2\pi s m,$ and this was calculated in \cite{dennis:2004twirl} to be $-sm/a.$
The loop has edge nature since, lying in the $xy$-plane, it is perpendicular to the global propagation in $z;$ its edge component is $k\left(1-(1+m)/2(1+m-a k)^2\right)$ in the $z$-direction.
More complicated threaded and linked vortex lines have been seen to occur generically in computer experiments of random optical speckle fields \cite{odp:2009told}, and knotted and linked, threaded vortices in holographically-controlled laser fields \cite{ldcp:2004knotted}.

The general relation between vortex linking and knotting and helicity was described by Moffatt \cite{moffatt:1969knottedness}, and the discussion there for $\delta$-concentrations of vorticity applies directly to the case of complex scalar wavefields.
Helicity has previously been discussed along similar lines to here for wave dislocations and optical vortices \cite{dzj:1999framework,rzd:2008knotted}.

There are two simple cases of optical waves which are often described to contain `screw dislocations': paraxial beams (satisfying $\nabla_{\bot}^2 \psi_{\rm{p}} = -2\rmi k \partial_z \psi_{\rm{p}}$), such as Laguerre-Gauss modes \cite{apb:1999oam}; such waves represent an approximation to paraxially small deviation from an exact plane wave travelling in the $z$-direction $\exp(\rmi k z) \psi_{\rm{p}}.$
In the paraxial approximation, $k$ is assumed to be infinitely larger than the largest transverse Fourier component in $\psi_{\rm{p}}$, so all the dislocation lines are approximately parallel to the $z$-direction.
Nonparaxial plane waves may also have a constant $z$-component $\exp(\rmi k_z z) \psi_{\rm{nd}},$ and these are the nondiffracting beams whose intensity is unchanged on propagation in $z,$ such as the Bessel beams \cite{md:2005bessel}.
It is easy to see in both these cases that the screw component of the wavevector, as defined in section \ref{sec:burgers}, is the constant magnitude of the component of the $z$-component of the wavevector: $k_z$ for nondiffracting beams, and for paraxial waves, approximately $k.$

\section{The edge component}
\label{sec:edge}

The edge component of the Burgers vector is a vector defined for all optical vortices (possibly zero for pure screw lines).
For two-dimensional fields, such as the transverse plane of a (paraxial or nondiffracting) beam, this vector lies in the transverse plane, and apparently does not correspond to any previously described quantity associated with optical phase singularities.
Unless stated otherwise, `the Burgers vector' in this section refers to the edge component of the Burgers vector.

\begin{figure}
\begin{center}
\includegraphics*[width=10.5cm]{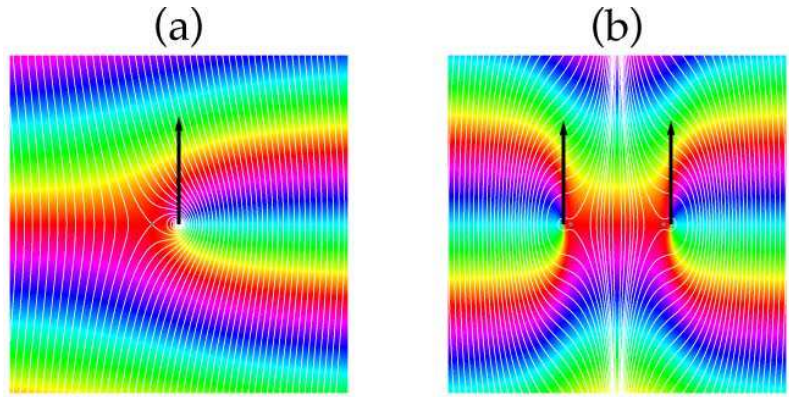}
\end{center}
    \caption{Edge dislocated waves in the plane. (a) $\psi_{k \hat{\bi{y}}}$ from Eq.~(\ref{eq:wavedislocation}), and (b) $\psi_{\rm{pair}}.$
    Colours denote phases labelled by hue, and the white curves show the streamlines of current and $\nabla \chi,$ which tend to follow the global wavevector in the $y$-direction.
    The Burgers vectors, also in the $y$-direction, are the black arrows, represented four times their true length.
    Saddle points are plotted in grey. }
    \label{fig:edge}
\end{figure}

The vector is the direction of the carrier wave local to the vortex line.
For two simple examples with a global carrier wave $\exp(\rmi k y),$ namely the edge dislocated wave $\psi_{k \hat{\bi{y}}}$ of (\ref{eq:wavedislocation}), and the exact two-dimensional solution of the Helmholtz equation \cite{nb:1974b34} $\psi_{\rm{pair}} = (k(x^2 - a^2) + \rmi y) \exp(\rmi k y),$ are shown in Fig.~\ref{fig:edge}.
The latter case has a pair of nodes with $s = \pm 1$ at $(\pm a, 0),$ annihilating when $a \to 0.$
Analytically, the Burgers vector of these two dislocations is $(0,k(1 - 1/(1 + 2 a k)^2)).$
Although in the limit $a \to 0,$ this is finite, the strength $s = 0$ at the annihilation event, and the assumptions in \ref{sec:contour} do not hold.
More general forms of annihilating dislocation points, as considered for instance in Ref.~\cite{nhh:1988tides}, have Burgers vectors which are the sum of any overall wavevector and a term related to the vortex anisotropy ellipses.
In three dimensions, the axisymmetric analogue of $\psi_{\rm{pair}}$ is an edge dislocation loop \cite{nb:1974b34}, perpendicular to the propagation direction, and may be realised by setting $m = 0$ in (\ref{eq:threaded}), for which the (purely edge) Burgers vector has length $k(1 - 1/2 /(-1 + a k)^2).$

In the two-dimensional edge plane perpendicular to the vortex line, there is a saddle point of the phase and flow of $\bi{j}$ (i.e.~$\chi$ finite, $\nabla \times \bi{j} < 0$), the Burgers vector is perpendicular to the displacement vector between the node and the saddle.
This phenomenon is straightforward to explain in the case of a dislocated field with a global wavevector, such as (\ref{eq:wavedislocation}); in the vicinity of the singularity ($R$ small), the main contributions to the vector field $\nabla \chi$ are the first two terms in (\ref{eq:gradchiexpand}): the circulation term, proportional to $R^{-1},$ and the Burgers vector term, constant when the wavevector is global.
Very close to the vortex, the circulating term dominates $\nabla \chi,$ but as $R$ increases, the constant Burgers vector term has more effect.
At the point in this neighbourhood when the two terms have equal magnitude and opposite direction, in direction $s(-b_y, b_x)$ with respect to $\bi{b} = (b_x,b_y)$ (assuming pure circulation), $\nabla \chi = 0,$ i.e.~at the saddle point.
Although the Burgers vector term is not constant for more general fields, it is natural to assume that phase saddle points occurring near dislocations \cite{nhh:1988tides,freund:1995saddles} are related to the direction of the Burgers vector in similar ways.

\begin{figure}
\begin{center}
\includegraphics*[width=8cm]{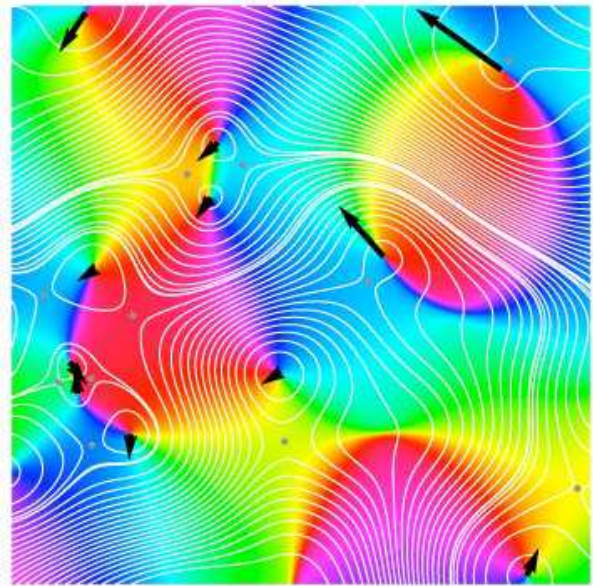}
\end{center}
    \caption{Four square wavelengths of an isotropic random wavefield, represented by the superposition of 100 plane waves with random directions, amplitudes and phases.
    The quantities represented are the same as in Fig.~\ref{fig:edge}.}
    \label{fig:random}
\end{figure}

A representation of a more complicated planar wavefield, which does not have a globally defined wavevector, is shown in Fig.~\ref{fig:random}.
The field depicted is a superposition of 100 plane waves with constant wavenumber $k,$ propagating in uniformly random directions with complex circular gaussian random amplitudes.
As is well known, there are many dislocations and saddle points (each with density $k^2/4\pi$ \cite{bd:2000b321}).
Also plotted are the stream lines of current and $\nabla \chi,$ and the (edge) Burgers vectors at the dislocations.
Most of the dislocations in the figure have a nearby saddle point, situated to the left of the dislocation with respect to the Burgers vector, as suggested by the preceding argument.
As can be seen on the upper right hand side of the figure, when there is a strong sense of direction to the current near isolated dislocations, the Burgers vector on the dislocation has a greater magnitude, and tends to be in this direction.
When the nearby current does not have a strong consensus direction, as in the lower right half of the figure, the Burgers vectors are smaller.
It is difficult to provide a more quantitative description of these phenomena in plane wave superpositions due to the form of $\bi{b},$ which is very complicated even in the simple case of superpositions of three plane waves.

\begin{figure}
\begin{center}
\includegraphics*[width=12cm]{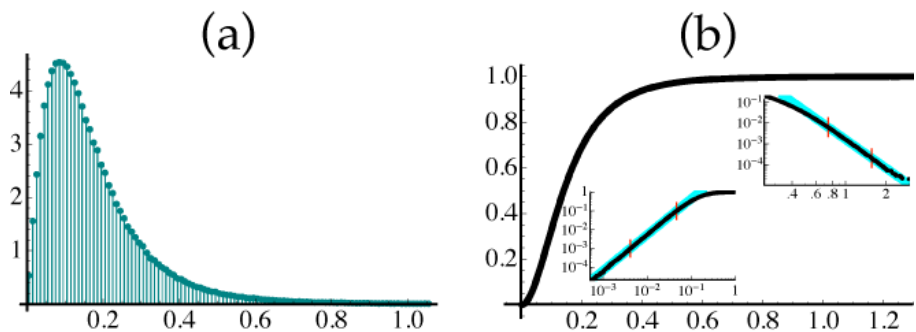}
\end{center}
    \caption{Probability of the Burgers vector magnitude $b$ for two-dimensional isotropic monochromatic fields. 
    (a) Probability density $P(b)$ represented by histogram (with horizontal axis units of $k$).
    (b) Cumulative probability $C(b),$ with insets showing log-log plots of $C(b)$ for small $b$ (lower left) and $1-C(b)$ for larger $b$ (upper right).
    The linear fits correspond to $35.9 b^{1.9}$ for small $b,$ and $.0016 b^{-4.57}$ for $b \gtrsim 0.8.$
    $b$ was calculated here from over 500 000 different dislocation points from large areas of several different realizations of superpositions of 200 random waves.}
    \label{fig:stats}
\end{figure}

A natural question to ask is what the statistical distribution of the Burgers vector $\bi{b}$ is for random waves of the type shown in Fig.~\ref{fig:random}, given the wealth of knowledge of other statistical features of wave dislocations (e.g.~\cite{bd:2000b321,freund:1994densities,dop:2009pio}).
In particular, the probability density function of the screw component of the Burgers vector for isotropic random three-dimensional waves was computed in \cite{dennis:2004twirl} (Eq.~(4.1)) to be a generalized Cauchy-type distribution.
However, the complicated form of the edge part of $\bi{b}$ has so far proved intractable to statistical methods.
The results of numerical computation of the length $b = \sqrt{b_x^2+b_y^2}$ of the Burgers vector for isotropic 2-dimensional monochromatic waves (or random nondiffracting waves in three dimensions) is shown in Fig.~\ref{fig:stats} as a histogram for the probability density function $P(b)$ in (a), and the cumulative probability distribution $C(b) \equiv \int_0^b P(b') \rmd b'$ in (b).
From this numerical data, it is possible to estimate the root mean square Burgers vector length $(\overline{b^2})^{1/2}$ as 0.22; clearly, the expected Burgers vector length is somewhat smaller than $k.$
Linear fits (shown in the insets to (b)) suggest that $P(b) \propto b$ for $b \approx 0,$ with power-law decay asymptotically, $P(b) \sim 0.007 b^{-5.5}$ for large $b.$
This suggests that there is a small but nonvanishing fraction of dislocations in this model for which the regularized local phase gradient $\bi{b}$ is larger than the global wavenumber $k:$ numerically, this fraction is approximately $0.0016.$
This contrasts with the fraction of nonsingular points whose local phase gradient magnitude exceeds $k,$ which is $1/3$ \cite{dhc:2008superoscillation}.

\section{Discussion}
\label{sec:disc}

All generic nodes/phase singularities in wave fields are vortices of energy, regardless of their specific geometry or orientation with respect to propagation direction.
However, other properties, such as the well-known helical wavefronts of optical screw dislocations, depend on the relationship of the zero with the global and local phase gradient, and this is summarised mathematically by the local phase gradient, here equated with the Burgers vector of the wave dislocation, in the spirit of Nye and Berry's approach.

This local phase gradient, represented in equations (\ref{eq:burgers}), depends both the first and second derivatives of the field at the node.
As such, it is a second order property of phase singularities, whereas quantities such as the vorticity (\ref{eq:S3def}) and other Stokes-like anisotropy parameters (\ref{eq:stokesdef}) only involve first derivatives.
Nevertheless, the Burgers vector plays an important role: topologically, in that its screw component, determining the local helicoidal twisting of the wavefronts, is related to the helicity of the wavefield; and physically, in that the edge component is a measure of the strength of current flow beyond the immediate vortical circulation.

This phase gradient is not the only quantity going beyond the immediate neighbourhood of optical vortices; for instance, the current streamlines themselves have interesting spiralling properties beyond first order expansions near vortices \cite{berry:2005b382,bl:2009nature}.
However, the spiralling defined in Ref.~\cite{berry:2005b382} depends on third as well as second derivatives of the field close to the vortex, and so is not immediately related to the quantities discussed here. 

Only basic properties of the Burgers vector have been discussed here, and further examination might yield further insights into optical singularity physics and the relationship between local and global energy flow near vortices.
For instance, can the edge component of the Burgers vector twist along wave dislocation lines? 
If so, does this define a higher-order helicity for closed loops?

The Burgers vector here defined might be related to the various phase gradient methods (e.g.~\cite{rls:1997dynamics,cr:2008accelerating}) proposed for the propagation dynamics of vortex points in paraxially propagating fields, which account for the transverse path of vortex points (effectively the tangent direction in three dimensions) by constructing a local phase gradient. 
Such a direction is either perpendicular (`glide') or parallel (`climb') to the direction of the Burgers vector (Nye and Berry \cite{nb:1974b34} studied such phenomena in time-dependent fields), and such a comparison with the crystallographic case may provide new understanding of the propagation of optical singularities.

\section*{Acknowledgements}

I am grateful to Anton Desyatnikov, John Hannay, Robert King and John Nye for illuminating discussions, and particularly thankful to Michael Berry for a crucial simplifying suggestion.
Part of this work was done during a visit to the Nonlinear Physics Centre at the Australian National University in Canberra, and I gratefully acknowledge the hospitality of Anton Desyatnikov and Yuri Kivshar.
This work was supported by the EPSRC, the Leverhulme Trust and the Royal Society of London.

\appendix

\section{Contour integration of the regularized phase gradient}
\label{sec:contour}

Integrals over $0 \le \phi \le 2\pi$ of functions such as 
\begin{eqnarray}
   f(\phi) = \frac{c}{a \cos \phi + b \sin \phi}, \label{fdef} \\
   g(\phi) = \frac{c \cos \phi + d \sin \phi}{a \cos \phi + b \sin \phi}, \label{gdef} \\  
   h(\phi) = \frac{c \cos^2 \phi + d \sin^2 \phi+ 2 e \cos \phi \sin \phi}{(a \cos \phi + b \sin \phi)^2},
   \label{eq:hdef}
\end{eqnarray}
with $a, b \neq 0,$ may be evaluated using complex contour integration methods.

In the following, $g(\phi)$ will be integrated as an example.
In doing so, it is useful to introduce the following function of complex variable $Z,$
\begin{equation}
   G(Z) \equiv \frac{c(Z+1/Z)- \rmi d (Z - 1/Z)}{a (Z+1/Z)- \rmi b (Z - 1/Z)} = \frac{Z^2 (c - \rmi d) + (c + \rmi d)}{Z^2 (a - \rmi b) + (a + \rmi b)},
   \label{eq:Gdef}
\end{equation}
since $G(\rme^{\rmi \phi}) = g(\phi).$
Around the unit circle $C$ in the complex $Z$-plane, $\rmd \theta = -\rmi\, \rmd Z / Z,$ so
\begin{equation}
   \int_0^{2\pi} \rmd \phi \, g(\phi) = -\rmi \oint_C \frac{\rmd Z}{Z} G(Z).
   \label{eq:gGint}
\end{equation}

$G(Z)$ is a meromorphic function of $Z,$ with simple poles at $Z_{\pm} = \pm \rmi \sqrt{(a + \rmi b)/(a -\rmi b)}.$
These poles are inside the integration contour if the sign
\begin{equation}
   S = \sign(| a + \rmi b|^2 - |a - \rmi b|^2) = \sign \im a^* b
   \label{eq:Sdef}
\end{equation}
is positive; when $g(\phi)$ corresponds to the $R$-independent term in (\ref{eq:bintdef}), $S = \sign \im \psi_x^* \psi_y = s,$ the singularity strength.
In this paper, it is always assumed that this is nonzero; otherwise there are poles on the integration path in the complex plane, and higher-order regularization of the phase singularity is necessary.

If $S = -1,$ then neither pole of $G(Z)$ is enclosed by the contour, and the only contribution to the integral (\ref{eq:gGint}) is the residue of the simple pole at the origin, giving
\begin{equation}
   \int_0^{2\pi} \rmd \phi \, g(\phi) = -\rmi \times 2 \pi \rmi G(0) = 2\pi \frac{c + \rmi d}{a + \rmi b} \qquad \hbox{(for $S < 0$)}.
   \label{eq:gint-}
\end{equation}

When $S = +1,$ the conjugate counterpart $\overline{G}(Z^*)$ to $G(Z)$ can be defined,
\begin{equation}
   \overline{G}(Z^*) \equiv \frac{Z^{*2} (c + \rmi d) + (c - \rmi d)}{Z^{*2} (a + \rmi b) + (a - \rmi b)},
   \label{eq:Fdef}
\end{equation}
with $g(\phi) = \overline{G}(\rme^{-\rmi \phi}),$ and on the unit circle in the $Z^*$-plane, $\rmd \phi = \rmi\, \rmd Z^*/Z^*,$ and the integration contour for the integral corresponding to (\ref{eq:gGint}) is clockwise.

The simple poles of $\overline{G}(Z^*),$ considered as a meromorphic function of $Z^*,$ are outside the unit circle, and do not contribute to the complex contour integral, which depends on the residue at the pole at $Z^* = 0$ in the usual way.
Thus, for any $g(\phi),$
\begin{equation}
   \int_0^{2\pi}\rmd \phi \, g(\phi) = 2\pi \frac{c - \rmi S d}{a - \rmi S b}, \qquad S = \sign \im a^* b.
   \label{eq:gint}
\end{equation}

By similar arguments, the meromorphic functions $F(Z), \overline{F}(Z^*)$ corresponding to $f(\phi)$ have poles in the same places; however, the integrand has no pole at the origin, and so
\begin{equation}
   \int_0^{2\pi}\rmd \phi \, f(\phi) = 0.
   \label{eq:fint}
\end{equation}

The integration of $h(\phi)$ follows a similar argument, and the corresponding meromorphic functions $H(Z), \overline{H}(Z^*)$ with $h(\phi) = H(\rme^{\rmi \phi}) = \overline{H}(\rme^{-\rmi \phi})$ now have double poles at the same places as the corresponding functions $H(Z), \overline{H}(Z^*).$
The arguments to integrate these are identical to those above, giving
\begin{equation}
   \int_0^{2\pi}\rmd \phi \, h(\phi) = 2\pi \frac{c - d  -2 \rmi S e}{(a - \rmi S b)^2}, \qquad S = \sign \im a^* b.
   \label{eq:hint}
\end{equation}

\section*{References}

\bibliographystyle{unsrt}

\end{document}